\documentclass[lettersize,journal]{IEEEtran}
\usepackage{amsmath,amsfonts}
\usepackage{algorithmic}
\usepackage{array}
\usepackage[caption=false,font=normalsize,labelfont=sf,textfont=sf]{subfig}
\usepackage{textcomp}
\usepackage{stfloats}
\usepackage{url}
\usepackage{verbatim}
\usepackage{graphicx}
\usepackage{cite}

\usepackage{orcidlink}

\newboolean{showcomments}
\setboolean{showcomments}{true}
\ifthenelse{\boolean{showcomments}}
 { \newcommand{\mynote}[2]{
      \fbox{\bfseries\sffamily\scriptsize#1}
        $[$#2$]$}}
{ \newcommand{\mynote}[2]{}
}

\ifthenelse{\boolean{showcomments}}
 { 
 }
{ 

}

\newcommand{\toolName}{RevMate}

\usepackage[linesnumbered,vlined,ruled,boxed,commentsnumbered]{algorithm2e}
\usepackage{xcolor}
\usepackage{tabularx}
\usepackage{diagbox}
\usepackage{multirow}
\usepackage{hyperref}

\usepackage{fancybox}
\usepackage{soul}

\SetKwInOut{Input}{Input}
\SetKwInOut{Output}{Output}
\SetKwProg{Throw}{throw}{}{}

\newcommand{\suggRQone}{How frequently do reviewers accept comments generated by an LLM-based approach?}
\newcommand{\suggRQtwo}{How does the comment category correlate with acceptance ratios?}
\newcommand{\suggRQthree}{How does the adoption of LLM-generated comments impact the code review workflow?}
\newcommand{\suggRQfour}{How does the adoption of LLM-generated comments impact the patch's review process?}

\usepackage{listings}
\usepackage{xcolor}

\definecolor{codegreen}{rgb}{0,0.6,0}
\definecolor{codegray}{rgb}{0.5,0.5,0.5}
\definecolor{codepurple}{rgb}{0.58,0,0.82}
\definecolor{backcolour}{rgb}{0.95,0.95,0.92}

\lstdefinestyle{mystyle}{
    backgroundcolor=\color{backcolour},   
    commentstyle=\color{codegreen},
    keywordstyle=\color{magenta},
    numberstyle=\tiny\color{codegray},
    stringstyle=\color{codepurple},
    basicstyle=\ttfamily\footnotesize,
    breakatwhitespace=false,         
    breaklines=true,                 
    captionpos=b,                    
    keepspaces=true,                 
    numbers=left,                    
    numbersep=5pt,                  
    showspaces=false,                
    showstringspaces=false,
    showtabs=false,                  
    tabsize=2
}
\begin{document}

\title{Impact of LLM-based Review Comment Generation in Practice:\\A Mixed Open-/Closed-source User Study}

\author{Doriane Olewicki~\orcidlink{0000-0002-3285-1884}, 
    Leuson Da Silva~\orcidlink{0000-0002-9086-9038}, 
    Suhaib Mujahid~\orcidlink{0000-0003-2738-1387}, 
    Arezou Amini~\orcidlink{0009-0006-6930-6587}, 
    Benjamin Mah~\orcidlink{0009-0003-9885-3799},\\ 
    Marco Castelluccio~\orcidlink{0000-0002-3285-5121}, 
    Sarra Habchi~\orcidlink{0000-0002-5989-1413}, 
    Foutse Khomh~\orcidlink{0000-0002-5704-4173}, 
    Bram Adams~\orcidlink{0000-0001-7213-4006}
}




\maketitle

\begin{abstract}
Code review is a standard practice in modern software development, aiming to improve code quality and facilitate knowledge exchange among developers. 
As providing constructive reviews on a submitted patch is a challenging and error-prone task, the advances of Large Language Models (LLMs) in performing natural language processing (NLP) tasks in a human-like fashion have prompted researchers to evaluate the LLMs' ability to automate the code review process.
However, outside lab settings, it is still unclear how receptive reviewers are to accept comments generated by LLMs in a real development workflow. 
To fill this gap, we conduct a large-scale empirical user study in a live setup to evaluate the acceptance of LLM-generated comments and their impact on the review process. 
This user study was performed in two organizations, Mozilla (which has its codebase available as open-source) and Ubisoft (fully closed-source). 
Inside their usual review environment, participants were given access to \toolName, an LLM-based assistive tool suggesting generated review comments using an off-the-shelf LLM with Retrieval-Augmented Generation to provide extra code and review context, combined with LLM-as-a-Judge, to auto-evaluate the generated comments and discard irrelevant cases. 
Based on more than 587 patch reviews provided by \toolName, we observed that 8.1\% and 7.2\%, respectively, of LLM-generated comments were accepted by reviewers in each organization, while 14.6\% and 20.5\% other comments were still marked as valuable as review or development tips. 
Refactoring-related comments are more likely to be accepted than Functional comments  (18.2\% and 18.6\% compared to 4.8\% and 5.2\%).
The extra time spent by reviewers to inspect generated comments or edit accepted ones (36/119), yielding an overall median of 43s per patch, is reasonable.
The accepted generated comments are as likely to yield future revisions of the revised patch as human-written comments (74\% vs 73\% at chunk-level).

\end{abstract}


\section{Introduction}
\label{sec:introduction}

Software code review is a core practice for modern software quality assurance~\cite{morales2015code,mcintosh2016empirical}, widely adopted in industrial and open-source projects ~\cite{sadowski2018modern,beller2014modern}.
While, initially, code review mostly was a synonym for code inspections on a submitted patch (structured as a set of chunks, i.e., successive lines of modified code)~\cite{bosu2017process,wang2015comparative},
the field gradually adopted a more dynamic approach to performing reviews commonly known as \emph{modern code reviews}~\cite{bacchelli2013expectations}, embracing social dimensions, like facilitating knowledge transfer between developers and strengthening synergy within teams~\cite{pangsakulyanont2014assessing}.

Despite such benefits, code reviews can also bring additional costs, due to the delay between a patch submission and its final approval for integration by the reviewers caused by back-and-forth between its author and reviewers~\cite{pascarella2018information}.
Additionally, providing valid and effective reviews requires non-trivial efforts from reviewers in terms of technical, social, and personal aspects~\cite{ebert2021exploratory}.
Reviewers need to understand and rationalize the overall impact of the changes under analysis while using effective communication~\cite{wurzel2023competencies,chouchen2021anti,paul2021security}.
Reviewers, even when qualified and focused on a patch, might miss issues due to fatigue~\cite{baum2019associating}, distraction, or pressure from other deadlines~\cite{shahin2017continuous}.

To overcome these limitations, the research community has explored different approaches to generate review comments with LLM-based approaches such as T5, CodeT5 and ChatGPT\cite{tufano2022using, li2022automating, tufano2024code}. However, thus far these have only been evaluated using textual metrics such as BLEU score and Exact Match, which are imperfect as they do not capture the actual impact and synergy the approach can have on reviewers~\cite{tufano2022using}. Furthermore, it is unclear how reviewers receive LLM-generated comments in the long-term, i.e., apart from their relevance, to what extent do generated comments solicit follow-up comments by the patch author or lead to follow-up patch revisions?

To fill in this gap, we conducted the first large-scale user study in a live setup involving two organizations, Mozilla and Ubisoft. 
Our choice for both an open-source (Mozilla) and closed-source (Ubisoft) organization is deliberate, since open-source codebases risk being 
part of the pretraining data of LLMs like GPT4o, yielding an unfair advantage in empirical studies.
Hence, this study also aims at investigating whether open-source organizations actually benefit from this potential advantage compared to closed-source organizations.

Spanning over 6 weeks, our user study involved 59 reviewers, covered 587 patch reviews, and led to the evaluation of 1.6k generated comments. 
During this study, we monitored the reviewers' interactions with the generated comments, assessed the impact of these comments on the review flow, and finally conducted a survey of the participating reviewers to gain insight into their perspectives, filled by 37/59 participants. 

For this evaluation, we built \textit{\toolName}, an LLM-based review assistant that generates review comments and that is easy to integrate into modern review environments.
\toolName{} builds on GPT4o and uses both (i) Retrieval Augmented Generation (RAG)~\cite{zhao2024retrieval} to enclose relevant information and ground the model on the project under analysis, and (ii) LLM-as-a-Judge~\cite{zheng2023judging} to leverage LLMs' capacity to evaluate generated content and discard irrelevant review comments.
In particular, we exploit two different approaches for \toolName:

\textit{\toolName{} with extra code context (Code):} inspired by Zhou et al.'s approach for code generation~\cite{zhou2022docprompting}, the model can request a code retrieval tool to provide {function definitions} and additional {code lines} from the codebase under analysis.

\textit{\toolName{} with related comment examples (Example):} like Parvez et al.'s approach for code summarization~\cite{parvez2021retrieval}, for each patch, the model dynamically selects few-shot examples of review comments similar to its chunks, then provides them to the LLM.

This paper addresses the following research questions:

\textbf{RQ1: \suggRQone} 

    Based on the participants' feedback in both organizations, we found that 8.1\% and 7.2\% of comments are accepted by Mozilla and Ubisoft reviewers, respectively, with 23\% and 28.3\% other comments still appreciated as being helpful.
    Despite the organizations differences in data availability, both the acceptance and appreciation ratios were consistent among them, indicating that the results may generalize across closed- and open-source development processes.
    Across both organizations, 23/37 participants reported they would continue using \toolName\ at least sometimes outside of the study context.
    
\textbf{RQ2: \suggRQtwo}

    We found that, for Mozilla and Ubisoft respectively,
    79.2\% and 84.9\% of generated comments are classified as functional issues, and 15.8\% and 14.5\% as refactoring-related comments.
    Furthermore, we found that refactoring-related comments have higher acceptance ratios than functional comments (18.2\% and 18.6\% compared to 4.8\% and 5.2\%).

\textbf{RQ3: \suggRQthree}

    We find that, although reviews take longer due to the extra time needed to evaluate the generated comments, the time spent per generated comment is reasonable (43s as a median per patch). 
    This time is mostly spent on investigating the validity of the comment and on editing the comment when applicable, as 37/119 accepted comments were edited, with 25/37 cases being shortening the comment.

    

\textbf{RQ4: \suggRQfour} 
   
    This RQ assesses the extent to which accepted generated comments lead to changes in the codebase. 
    We found that accepted review comments lead to as many patch revisions (i.e., changes in a future version of the patch) as human comments (74\% vs 73\% at chunk level) in Ubisoft, indicating that the generated comments have the same impact as human comments.
    Also, generated comments lead to fewer follow-up comments among developers (23\% compared to 34\%).
    


The main contributions of this study are:
\begin{itemize}
    \item A large-scale mixed closed-/open-source user study, with 57 expert reviewers, assessing the usefulness of LLM-generated review comments and their impact on the review flow and outcome;
    \item A diverse set of 3.4k generated code review comments by GPT4o for the Mozilla organization, among which 426 were evaluated by reviewers~\cite{datamoz};
    \item An open source project implementing the comment review generation approach~\cite{bugbug}. 
\end{itemize}


\section{Related Work}
\label{sec:related-work}

\subsection{Automating the Code Review Process}
Previous studies have proposed a variety of approaches to automate different stages of the code review process.
For example, one category of studies focuses on recommending reviewers for a given patch~\cite{al2020workload,ouni2016search}.
Other studies investigate code change quality prediction, i.e., the best location in a submitted patch where review comments should be added~\cite{li2022automating,hellendoorn2021towards,shi2019automatic,hong2022should}.
Such approaches can assist the review process by either signaling that a chunk of code should be focused on~\cite{henley2018cfar}, or reordering files to show problematic parts of the patch first~\cite{olewicki2024empirical}.
However, they still rely on human effort to perform the review, which can be a time-consuming task~\cite{macleod2017code}.


\subsection{Review Comment Suggestion}

The field of review comment recommendation studies the generation of comments for a submitted patch.
Over the years, several such approaches have been proposed, which either represent code changes using tokenization~\cite{gupta2018intelligent, siow2020core}, word- and character-level embeddings~\cite{siow2020core}, or bag-of-words approaches~\cite{tufano2021towards, tufano2022using, li2022automating, hong2022commentfinder}. 
State-of-the-art (SOTA) approaches for review comment generation leverage custom LLMs such as T5 and CodeT5~\cite{li2022automating,tufano2022using}.
Tufano et al.~\cite{tufano2024code} compare these LLM-based SOTA approaches for code review generation to ChatGPT used in a zero-shot cognitive architecture, i.e., without any customization.
Although the authors report that ChatGPT's outputs often miss or contradict the main points raised by the human reviewer, the use of zero-shot prompting suggests that the use of additional context, e.g., using Retrieval Augmented Generation, could still improve on the off-the-shelf model's performance~\cite{zhao2024retrieval}, which we explore in this paper.

Hong et al.~\cite{hong2022commentfinder} propose \emph{CommentFinder}, a tool that retrieves past code review comments of similar patches and reports them as-is for the current patch under review.
They motivate their approach based on many other studies that found simpler approaches to be similar or even better than Deep Learning (DL) approaches in the context of software engineering (SE) tasks in terms of performance~\cite{fu2017easy,hellendoorn2017deep} and speed~\cite{majumder2018500}.
We believe that generative approaches could take advantage of \emph{CommentFinder}'s idea of selecting comments associated with the patch under review as few-shot examples in the prompt, which we considered in our study. 

Prior work~\cite{tufano2021towards,li2022automating,hong2022commentfinder} uses traditional NLP metrics, such as BLEU score (i.e., a metric of text translation quality) and Exact Match (i.e., the prediction matching exactly the expected output). 
While Tufano et al. report BLEU scores around 5\% and Exact Match of 3\%, they suggest that these metrics are not perfect, and might thus under-evaluate the model's performance~\cite{tufano2022using}. Hence, the need for either investigating the creation of better metrics or conducting user studies and manual validations by experts: while previous work considered the former~\cite{huang2024empirical}. While Unzicker et al. conducted a user study to evaluate through a survey the impact of LLMs on mental workload and performance regarding the review comment generation task~\cite{unzicker2024all}, we evaluate its acceptance and usage of generated comments in day-to-day reviews.


\subsection{Leveraging LLMs and Prompt Engineering}


Amidst the many LLMs currently available, GPT, a family of LLMs developed by OpenAI, currently stands out as one of the most popular LLMs. 
GPT has been used for a variety of applications, among which SE use cases like code generation, summarization, testing, and documentation \cite{zheng2023towards, ozkaya2023application}.

To guide LLMs like GPT in performing specific tasks accurately, users make requests to LLMs through textual prompts.
This process of defining, maintaining, and improving prompts is the subject of Prompt Engineering, a field that combines artificial intelligence and NLP tasks~\cite{beurer2023prompting}.
Various prompting patterns have been proposed to structure prompts more effectively.
For instance, while \emph{zero-shot} prompts rely exclusively on textual instructions given directly to an LLM~\cite{radford2019language}, \emph{few-shot} approaches provide more explicit context to the LLM by including concrete examples provided by the LLM user, to guide the LLM towards the expected behavior~\cite{nashid2023retrieval,brown2020language}. 
\emph{Chain-of-Thought} consists of providing a set of intermediate reasoning steps in the textual instructions, supporting the LLM to address tasks that require complex reasoning~\cite{wei2022chain}.

To improve the quality of a model's output, several techniques have been proposed. For instance, a user's prompt may specify constraints in terms of the style, structure, or type of answer expected from the model.
Among those, the prompt can suggest intrinsic characteristics like human pre-defined behavior patterns or persona that the LLM should inpersonate~\cite{white2023prompt}.
Also, \emph{Retrieval-Augmented Generation} can be used to
enhance the prompt with extra information relevant to the prompt retrieved from a corpus of user-provided documents, before the enriched prompt is sent to the LLM~\cite{zhao2024retrieval}. For instance, RAG has been used in SE for code generation~\cite{zhou2022docprompting} and code summarization~\cite{parvez2021retrieval}.
Finally, \textit{LLM-as-a-Judge} leverages the capacity of an LLM to auto-evaluate its generated content, which can be used to improve review comment generation through filtering or iterative generation~\cite{zheng2023judging}.
Our review comment generation approach integrates those different advances in LLM prompting.

\section{User Study Design}
\label{sec:methodology}

Our goal is to evaluate the impact of integrating review comment generation into reviewers' workflow using a large-scale, mixed open-/closed-source user study. To do so, we developed a review assistant (\textit{\toolName}) that integrates easily into common review environments.
This section discusses the approaches used by \toolName\ and how the feedback of reviewers is gathered from the tool in the context of a user study.

\subsection{\toolName's LLM-based approach}
\label{sec:workflow}
\toolName{} relies on the following common LLM-based mechanisms: RAG (tool-LLM collaboration for code retrieval) and LLM-as-a-Judge through LLM-based evaluation. 
The approach leverage the Chain-of-Thought architecture, by splitting the overall review comment generation process 
into multiple prompts, some of which (i.e., prompts (5) and (6) in Table~\ref{tab:prompts}) also use Chain-of-Thought internally. Furthermore, a memory conversation is used to tie the prompts together ((4) in Table~\ref{tab:prompts}). 
In the following, we will describe the resulting workflow, summarized in Algorithm~\ref{alg:tool-workflow}.

\SetKwInput{KwData}{Input}
\SetKwInput{KwResult}{Output}

\begin{algorithm}[t]
    \caption{\toolName\ Workflow. The (x) notation refers to the corresponding prompt in Table~\ref{tab:prompts}.}
    \label{alg:tool-workflow}
    {\fontsize{9pt}{9pt}\selectfont
    \SetAlgoLined
    \KwData{$patch$, $approach$}

    \KwResult{$code\_review$}
    

    \uIf{$patchNeedsReview(patch.status)$}{
                
                $fpatch \leftarrow format(patch.sourcecode)$
                
                $sum \leftarrow askLLMforSummary(fpatch)$ (1)

                $mem \leftarrow createMemory([sum ])$ (4)

                \uIf{approach = Code}{
                $funcs \leftarrow askLLMNeedsFuncs(fpatch)$ (2)

                $lines \leftarrow askLLMNeedsLines(fpatch)$ (3)
                
                
                \uIf{$funcs.\text{size}$ $>$ 0}{
                    $context\_funcs \leftarrow getContext(funcs)$
                    
                    $mem.\text{addToMemory}(context\_funcs)$ (4)
                }

                \uIf{$lines.\text{size}$ $>$ 0}{
                    $context\_lines \leftarrow getContext(lines)$
                
                    $mem.\text{addToMemory}(context\_lines)$ (4)
                
                }
                
                $ex \leftarrow default\_set\_of\_few\_shot\_examples$
                
                }      

                \uElseIf{approach = Example}{
                $ex \leftarrow getSimilarExamples(fpatch)$
                }
                $first\_review \leftarrow askLLMReview(fpatch, mem,$ $ex)$ (5)
                
                $code\_review = askLLMFilterGeneration(fpatch,$
                $first\_review)$ (6)

                \Return $code\_review$
        }
    }
\end{algorithm}

\begin{table*}[t]
    \centering
    \caption{Prompts and conversation used by \toolName}
    \label{tab:prompts}
    \begin{tabularx}{\textwidth}{|X|}
    \hline
\textbf{(1) Summary prompt (\textit{askLLMforSummary}):} \\\hline
    You are an expert reviewer for source code, with experience in source code reviews.
    Please, analyze the code provided and report a summarization about the new changes; for that, focus on the code added represented by lines that start with "+". \newline
    \textit{\{patch\}}\\
    \hline
    
\textbf{(2) Function request prompt (\textit{askLLMNeedsFuncs}):} \\\hline
Based on the patch provided below and its related summarization, identify the functions you need to examine for reviewing the patch.
List the names of these functions, providing only the function names, with each name on a separate line.
Avoid using list indicators such as hyphens or numbers.
If no function declaration is required, just return "".\newline
\textit{\{patch\}}
\textit{\{summarization\}} \\ \hline

\textbf{(3) Line context request prompt (\textit{askLLMNeedsLines}):} \\\hline
Based on the patch provided below and its related summarization, report the code lines for which more context is required.
For that, list the lines with their associated line numbers, grouping each one on a separate line.
Avoid using list indicators such as hyphens or numbers. If no code line is required, just return "".
Examples of valid code lines: '152    const selector = notification.getDescription();' and '56        file.getElement(this.targetElement());'\newline
\textit{\{patch\}}
\textit{\{summarization\}}\\\hline

\textbf{(4) Memory conversation content (\textit{memory}):} \\ \hline

\textbf{in:} You are an expert reviewer for source code, with experience on source code reviews.\newline 
\textbf{out:} Sure, I'm aware of source code practices in the development community. \newline
\textbf{in:} Please, analyze the code provided and report a summarization about the new changes; for that, focus on the code added represented by lines that start with "+". \textit{\{patch\}}\newline
\textbf{out:} \textit{\{summarization\}}\newline
if requested context: \newline 
\textbf{in:} Attached, you can find some function definitions that are used in the current patch and might be useful to you, by giving more context about the code under analysis.
\textit{\{function context\}}\newline
\textbf{out:} Okay, I will consider the provided function definitions as additional context to the given patch.\newline
\textbf{in:} Attached, you can also have more context of the target code under analysis.
\textit{\{line context\}}\newline
\textbf{out:} Okay, I will also consider the code as additional context to the given patch.
\\\hline

\textbf{(5) Review generation prompt (\textit{askLLMReview}):} \\\hline
    You will be given a task to generate a code review for the patch below. Use the following steps to solve it:\newline
    1. Understand the changes done in the patch by reasoning about the summarization as previously reported.\newline
    2. Identify possible code snippets that might result in possible bugs, major readability regressions, and similar concerns.\newline
    3. Reason about each identified problem to make sure they are valid. Keep in mind, your review must be consistent with the source code.\newline
    4. Filter out comments that focus on documentation, comments, error handling, tests, and confirmation whether objects, methods and files exist or not.\newline
    5. Filter out comments that are descriptive and filter out comments that are praising (example: "This is a good addition to the code.").\newline
    6. Filter out comments that are not about added lines (have a '+' symbol at the start of the line).\newline
    7. Final answer: Write down the comments and report them using the JSON format previously adopted for the valid comment examples.\newline
    As valid comments, consider the examples below:
    \textit{\{ex\}}
    \newline
    Here is the patch that we need you to review:
    \textit{\{patch\}}\\ \hline

\textbf{(6) Review generation filtering prompt (\textit{askLLMFilter}):} \\\hline

Please, double check the code review provided for the patch below.
Just report the comments that are:\newline
- applicable for the patch;\newline
- consistent with the source code;\newline
- focusing on reporting possible bugs, major readability regressions, or similar concerns;\newline
- filter out any descriptive comments;\newline
- filter out any praising comments;\newline
- filter out any comments that do not suggest code changes.\newline
Do not change the contents of the comments and the report format.
Adopt the template below as the report format:\newline
[
    \{
        "file": "com/br/main/Pressure.java",
        "code\_line": 458,
        "comment" : "a generated comment"
    \}
]\newline
Do not report any explanation about your choice.\newline
Review:
\textit{\{review\}} -
Patch:
\textit{\{patch\}}\newline
As examples of unexpected comments, unrelated to the current patch, please, check some below: \textit{\{undesired\_comments\}}\\ \hline
    \end{tabularx}
\end{table*}

\subsubsection{Starting Point}
\toolName\ is automatically called when a reviewer opens a patch in their review environment.
Only patches labelled as \emph{Needs Review} are analyzed (line 1, Algorithm \ref{alg:tool-workflow}). 
Review comments are generated once per patch and kept in cache to maintain consistency if reviewers reload the page. 
Moreover, a reviewer will only be able to see any previously generated comments that have not yet been evaluated by another reviewer.

\toolName\ re-formats the patch to help the LLM understand the patch's content (line 2, Algorithm \ref{alg:tool-workflow}).
Knowing that a patch consists of one or more chunks (i.e., a sequence of changed lines), \toolName\ groups a patch's chunks per file while also adding the corresponding file line number to the beginning of each code line (instead of Git's default line number-range per chunk).
This allows the LLM to better associate the generated review comments with the target lines.


Next, \toolName\ requests the LLM to summarize the formatted patch (line 3, Algorithm \ref{alg:tool-workflow}; prompt (1) in Table~\ref{tab:prompts}), using the LLM's ability to self-explain code~\cite{zheng2023take, wei2022chain,li2024approach}.

\subsubsection{Extra context}~\label{approach_sec}
After generating the summary, \toolName{} seeks extra context to ground the model on the codebase under analysis and improve the quality of the generated comments. 
Building on the RAG mechanism, we consider two ways of providing extra context for comment generation, i.e., \textbf{\textit{Code}} and \textbf{\textit{Example}}. Although both approaches are complementary, in our user study on the impact of generated comments on the review process, we only considered one approach at a time.

\paragraph{Code context (\textbf{Code})}
Being aware that the patch under analysis might not contain sufficient details to perform a full analysis, we ask the LLM whether there are function calls referred by the patch that require the full function definition to understand its use (line 6, Algorithm \ref{alg:tool-workflow}).
For that, we feed the patch and its associated summary to the model, and request a list of the needed functions' names (prompt (2), Table~\ref{tab:prompts}).
Then, for each function name, a code retrieval tool accesses the respective project repository to retrieve the function's definition~\cite{ardito2020rust} (line 9, Algorithm \ref{alg:tool-workflow}) to provide to the model via the memory (line 10, Algorithm \ref{alg:tool-workflow}; conversation (4), Table~\ref{tab:prompts}).

Besides functions' definitions, other code context can also be needed for the model's review generation. In particular, the extracted code chunks only show up to 10 lines of context (i.e., unmodified lines above or below modified lines). This too might lead to missing context and hinder patch understanding.
Thus, we also ask the LLM whether it requires additional context for specific patch lines (line 7, Algorithm \ref{alg:tool-workflow}; prompt (3), Table~\ref{tab:prompts}), this time getting extended context around the requested lines by retrieving the functions that include said lines (line 12, Algorithm \ref{alg:tool-workflow}), then providing it to the memory (line 13, Algorithm \ref{alg:tool-workflow}; conversation (4), Table~\ref{tab:prompts}).

Finally, when using the \textit{Code} approach, the comment generation prompt (prompt (5), Table~\ref{tab:prompts}) will be provided with a fixed list of few-shot examples $ex$ randomly selected from past examples of chunks with their associated review comments (line 14, Algorithm \ref{alg:tool-workflow}). This default set is available in the replication package~\cite{bugbug}.


\paragraph{Related comment examples (\textbf{Example})} 
The \textit{Example} approach improves on the default set of examples provided by the \textit{Code} approach, by explicitly selecting relevant examples chosen based on their similarity to the current patch (line 16, Algorithm \ref{alg:tool-workflow}).
To do so, we first constructed a dataset for each studied project comprising past chunks and their associated comments (tuples of $(chunk, com)$), excluding tuples where the comment had more than 500 characters (to encourage concise comments only), or comments that included urls (which the model does not have access to). The examples are loaded in a vector database~\cite{qdrant} with GPT embeddings. Examples are retrieved from this database using the cosine distance between the embeddings of the current patch's chunks and the historical ones stored in the vector database. We retrieve 10 examples for each chunk in the current patch, then report the 10 examples with highest similarity out of all retrieved examples for the patch.

\subsubsection{Generating code reviews}
The information retrieved (i.e., summary and extra context) is sequentially structured into a memory conversation buffer ((4) in Table~\ref{tab:prompts}), which is responsible for providing the task context, including code context for the \textit{Code} approach and relevant examples for the \textit{Example} approach.
First, we set up the LLM to assume the persona of an expert reviewer.
Second, we feed the patch summary, providing context for the changes under analysis (line 4, Algorithm \ref{alg:tool-workflow}).
If functions and context lines are provided by the code retrieval tool based on previous steps, they are also added to the buffer (lines 10\&13, Algorithm \ref{alg:tool-workflow}).

Finally, \toolName\ asks the model to perform a code review (line 17, Algorithm \ref{alg:tool-workflow}; prompt (5) in Table~\ref{tab:prompts}): we instruct it to follow a \emph{few-shot} approach by providing examples of valid comments, either the default set or the retrieved set, aiming to guide the LLM to provide related and relevant content~\cite{nashid2023retrieval,brown2020language}. 
This way, we impose a standard pattern for the comments, consisting of the comment, associated line number, and file ($\{com;line;file\}$).

\subsubsection{Filtering generated comments}
Once the LLM has generated the comments, \toolName\ makes one final LLM-request in another conversation (i.e., without memory) to filter out comments that do not address valid problems.
Leveraging the concept of LLM-as-a-Judge~\cite{zheng2023judging}, \toolName{} prompts a filtering model providing the list of generated comments and the associated patch as input (line 18, Algorithm~\ref{alg:tool-workflow}; prompt (6) shown in Table~\ref{tab:prompts}). The model is also provided with a list of undesired comment examples (\textit{undesired\_comments}, which we constructed from preliminary feedback from a early users at Mozilla and iterations with the prompt) to guide the filtering process, which are available in our replication package~\cite{bugbug}.
Last, \toolName\ guides the LLM to report its results in JSON format (i.e., list of $\{com;line;file\}$).

With the final list of comments generated (line 19, Algorithm~\ref{alg:tool-workflow}), \toolName\ stores them in a database server.
Based on the reported line numbers and file names associated with each comment, \toolName\ shows the generated comments directly on the patch under analysis in the review environment.

\subsubsection{Technical Details}

After initial experiments, we decided on
using GPT4o with a temperature of .2 to provide more accurate and deterministic results compared to higher temperatures, following Lin et al.~\cite{lin2024llm}.
For the embeddings used for the \textit{Example} approach, we used GPT text-embedding-3-large also with a .2 temperature. 
However, \toolName\ has been built for ease of extensibility and could be easily adapted to work with different LLMs and under different configurations.

\begin{figure}[t]
    \centering
    \includegraphics[width=.9\linewidth]{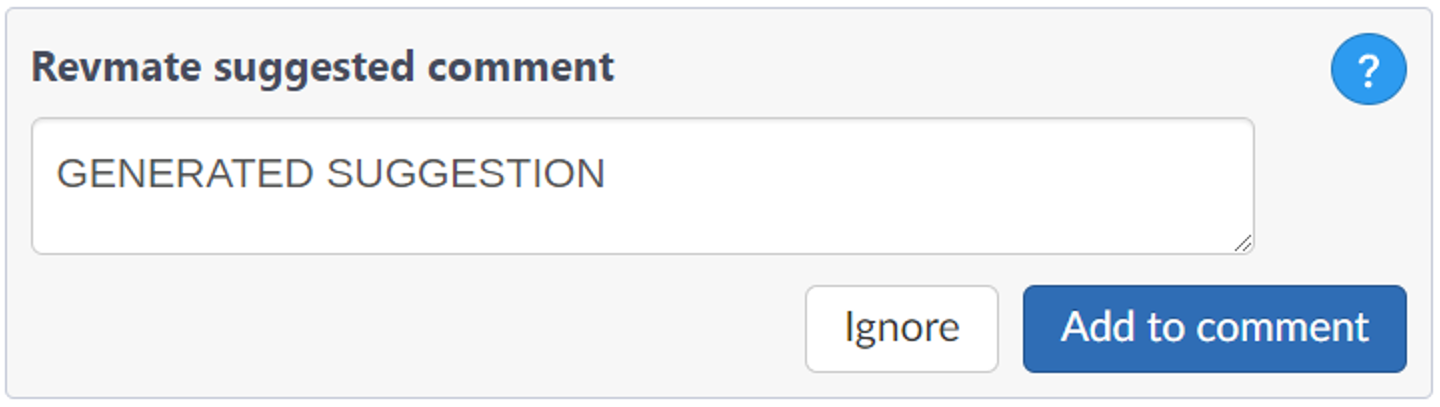}
    \caption{UI for generated comments' evaluation boxes}
    \label{fig:eval}
\end{figure}

\begin{figure}[t]   
    \centering 
    \includegraphics[width=.9\linewidth]{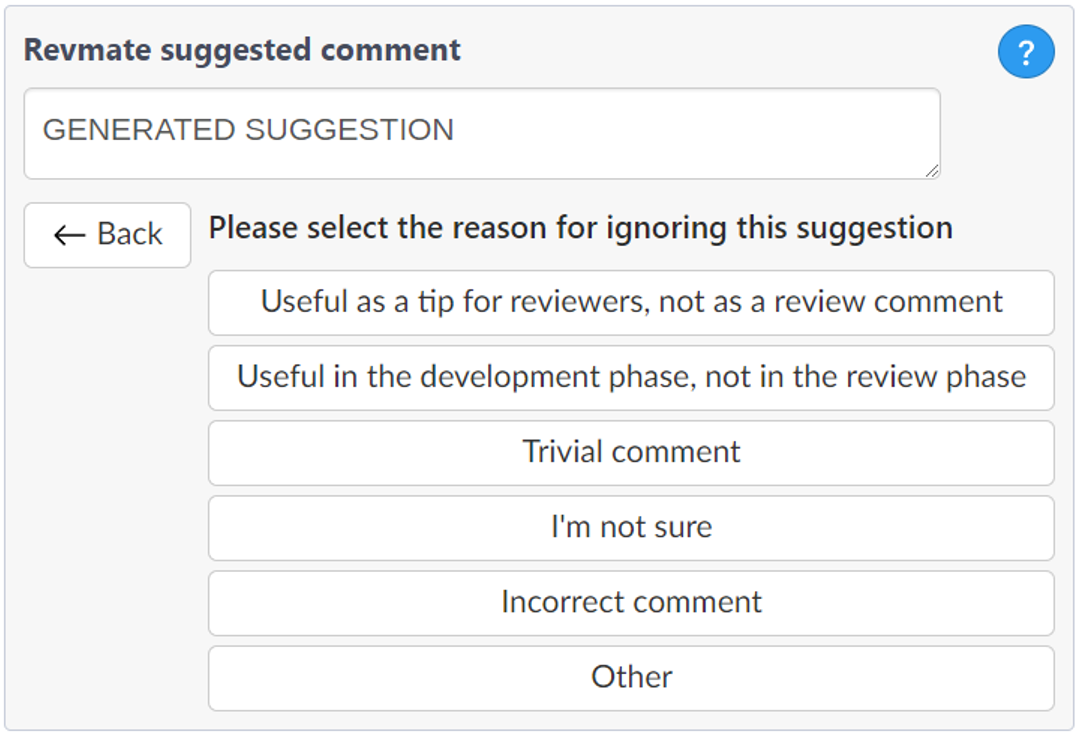}
    \caption{UI for choice of reasons for ignoring generated comments}
    \label{fig:ignore}
\end{figure}

\subsection{UI for \toolName's generated comments}
\label{sec:diffCompany}

we integrated \toolName\ into participants' review environment, making it available to them when performing their day-to-day code review tasks.
Through the user study, reviewers can evaluate each generated comment, which is presented as a suggestion through an evaluation box shown in Figure~\ref{fig:eval}: ``add to comment" to \emph{accept} it or ``ignore" to \emph{reject} it.
In case of rejection, we additionally asked reviewers to provide a reason for ignoring comments, giving the options shown in Figure~\ref{fig:ignore}.
Before evaluating the comment, reviewers can also edit its content. 
Once evaluated, the evaluation box disappears, leaving no trace in the case of rejection and leaving a publicly published comment in the case of acceptance.
Each generated comment can only be evaluated once.

As the participating organizations use two different review environments (Mozilla uses Phabricator\footnote{Phabricator, accessed: 2024-10-15. \url{secure.phabricator.com/}}, while Ubisoft uses Swarm\footnote{Swarm, accessed: 2024-10-15. \url{www.perforce.com/products/helix-swarm}}), we also had to adapt \toolName\ to follow each organization's specific policies regarding reviews and generative-AI. This led to the following differences in design choices: 

\textit{Mozilla:} Accepted generated review comments had to be published as a separate reviewer named ``\textit{\toolName}". 
The tool shows the generated comments' evaluation boxes directly when opening a review. 
In order to nudge the user study participants towards evaluating all review comments, the UI delays the publication of accepted generated comments to the patch author until all generated comments have been evaluated.


\textit{Ubisoft:} The reviewers had to publish the generated review comments that they accepted in their own name.
Furthermore, due to the review environment design and API, the generated comments could not be displayed immediately when opening the page. 
To mark where generated comments were, we added a star symbol next to the lines on which reviewers needed to click for it to appear. 
To further help reviewers with finding the generated comments, for each file, we indicated next to the filename the comments' line numbers. 

For both organizations, a summary at the beginning of the patch's page tells users how many generated comments were generated and need to be evaluated.

\subsection{User study setup}


We now discuss the setup of the user study that we designed to address the research questions listed in the introduction\footnote{Ethical approval: GREB Initial Ethics Clearance TRAQ\#6040278}.

\subsubsection{Participants and Target Projects} 
~\label{sec:companies}

Since our main goal is to evaluate the real-world potential of adopting LLMs as review assistant, we randomly contacted 50+ participants actively performing code reviews in ongoing projects in each organization. 
The targeted projects all are mature projects that are continuously updated and written in different programming languages (e.g., C++, C\#).

As the contacted users joined on a voluntary base, we did not have control on the demographics. However, we can report that Mozilla had 28 participants with an average experience of 8.5~years inside the organization (from 1-13~years) and 13.4~years in their career (from 5-25~years), while Ubisoft had 31 participants, with average experience of 7.6~years inside the organization (from 1-25~years) and 10.9~years in their career (from 4-25~years). 

Before starting the study, a manual with instructions to install and run the tool in their specific review environment was sent to them.

\subsubsection{Running the Study}

Participants of the study within each organization are randomly split into two groups: one group gets the \textit{Code} approach and the other gets the \textit{Example} approach (see Section~\ref{approach_sec}). The differences between the two approaches are studied in RQ1.

For practical reasons, the study was conducted during 5~weeks at Mozilla and 6 at Ubisoft, covering respectively 165 and 422 code reviews assisted by \toolName.
The set of all generated reviews from Mozilla can be found in our online Appendix~\cite{datamoz}, whereas Ubisoft's data has to remain closed-source.





\subsection{Data Collection}

For each generated review comment, we gathered the following information: comment identifier, line number, filename, patch identifier, reviewer identifier, timestamp of evaluation, content of the generated comment, edited content (if applicable), evaluation (i.e., accept or ignore), reason (choice from list in Figure~\ref{fig:ignore}).
In the case of Ubisoft, we also have the timestamp on which the generated comment is opened, as users had to open them to read then evaluate them, which we used for RQ3.

After having access to the tool for a few weeks, we sent a survey to users to get their general feedback on the generative tool. The questions can be found in Table~\ref{tab:survey}.
Mozilla received responses from  15/28 participants, while Ubisoft received responses from 22/31 participants, for a total of 37 responses.

\begin{table}[t]
    \centering
    \caption{End of study survey questions}
    \label{tab:survey}
    \begin{tabularx}{\linewidth}{|l|X|}
    \hline
       \#& \textbf{Question} \\
    \hline
    1&What is your experience as a developer in the company? (in years)\\\hline
    2&What is your experience as a developer in your carreer? (in years)\\\hline
    3&When using the tool, how was the duration of your reviews affected? \newline
    Slower - A bit slower - Same time - A bit quicker - Quicker
    \\\hline
    4&What would you change about the generated comments' content, if anything? (open) \\\hline
    5&Would you use a review tool like this tool during your review processes outside of the study? \newline 
    Never - Rarely - Sometimes - Often - Always\\\hline
    \end{tabularx}
\end{table}

\subsection{Metrics and statistical tests}


The accepted comments obviously represent high-quality generated comments, as they were adopted by reviewers. 
Hence, we analyze the acceptance ratio, $\frac{\#Accepted}{\#Evaluated}$, for both organizations.
However, among comments that were not accepted, some rejection motivations still indicate a positive impact for comments found to be useful as a tip for reviewers or during the development phase, which we will collectively refer to as ``Valuable Tips" ($\#ValuableTip$). We thus also report a second ratio, i.e., the appreciation ratio, $\frac{\#Accepted + \#ValuableTip}{\#Evaluated}$.

To compare proportions in our analysis, we use Fisher's statistical test, with $\alpha=.05$.
When the difference is statistically significant, we then use Cohen's d score ($CD$) to identify the effect~\cite{muller1989statistical}, with effect size small ($>$.2), medium ($>$.5), or large ($>$.8).


\section{RQ1: \suggRQone}

\begin{table*}[t]
    \centering
    \caption{RQ1 - Evaluation counts. ``Total" refers to the union of all the generated comments, disregarding the approach used.}
    \label{tab:rq1eval}
    \begin{tabularx}{\textwidth}{l|l||XX|X||XX|X}
    \hline
         \textbf{Macro-}& \multirow{3}{*}{\textbf{Evaluation}}& \multicolumn{3}{c||}{\textbf{Mozilla}}& \multicolumn{3}{c}{\textbf{Ubisoft}} \\
         \textbf{Evaluation}&&\multicolumn{2}{c|}{\textbf{Approach}}&\multirow{2}{*}{\textbf{Total}}&\multicolumn{2}{c|}{\textbf{Approach}}&\multirow{2}{*}{\textbf{Total}}\\
          &&Code&Example&&Code&Example&\\
         \hline

\hline \multirow{1}{*}{Accepted} & Accept & 9 (6.8\%) & 25 (8.5\%) & 34 (8.0\%) & 31 (5.3\%) & 54 (8.5\%) & 85 (7.0\%) \\
\hline \multirow{1}{*}{Valuable Tip} & Ignore - Useful as a tip for reviewers & 19 (14.4\%) & 20 (6.8\%) & 39 (9.2\%) & 51 (8.7\%) & 43 (6.8\%) & 94 (7.7\%) \\
 & Ignore - Useful for developer & 10 (7.6\%) & 13 (4.4\%) & 23 (5.4\%) & 105 (17.9\%) & 51 (8.1\%) & 156 (12.8\%) \\
\hline \multirow{1}{*}{Rejected} & Ignore - Trivial comment & 37 (28.0\%) & 114 (38.8\%) & 151 (35.4\%) & 112 (19.1\%) & 66 (10.4\%) & 178 (14.6\%) \\
 & Ignore - Incorrect comment & 42 (31.8\%) & 100 (34.0\%) & 142 (33.3\%) & 47 (8.0\%) & 137 (21.7\%) & 184 (15.1\%) \\
 & Ignore - Other & 10 (7.6\%) & 19 (6.5\%) & 29 (6.8\%) & 28 (4.8\%) & 50 (7.9\%) & 78 (6.4\%) \\
 & Ignore - Seen & / & / & / & 206 (35.2\%) & 204 (32.3\%) & 410 (33.7\%) \\
 \hline \textit{Filter out} & Ignore - I'm not sure & 5 (3.8\%) & 3 (1.0\%) & 8 (1.9\%) & 6 (1.0\%) & 27 (4.3\%) & 33 (2.7\%) \\
\hline \hline \multicolumn{2}{r||}{\textbf{Total \#comments}}
 & 132 & 294 & 426 & 586 & 632 & 1218 \\
 \multicolumn{2}{r||}{\textbf{Total \#evaluated}}
 & 127 & 291 & 418 & 580 & 605 & 1185 \\
\multicolumn{2}{r||}{\textbf{Total \#reviewers}}
 & 19 & 22 & 28 & 17 & 14 & 31 \\
\multicolumn{2}{r||}{\textbf{Total \#reviews}}
 & 59 & 106 & 165 & 197 & 227 & 422 \\
 
     \end{tabularx}
\end{table*}

\begin{table}[t]
    \centering
    \caption{RQ1- Evaluation ratios. ``Total" refers to the union of all the generated comments, disregarding the approach used.}
    \label{tab:rq1res}
    \begin{tabularx}{\linewidth}{l||XX|X||XX|X}
    \hline
         & \multicolumn{3}{c||}{\textbf{Mozilla}}& \multicolumn{3}{c}{\textbf{Ubisoft}} \\
         \textbf{Ratio}&\multicolumn{2}{c|}{\textbf{Approach}}&\multirow{2}{*}{\textbf{Total}}&\multicolumn{2}{c|}{\textbf{Approach}}&\multirow{2}{*}{\textbf{Total}}\\
          &Code&Ex&&Code&Ex&\\
         \hline


\hline 


Acceptance [\%] & 7.1 & 8.6 & 8.1 & 5.3 & 8.9 & 7.2 \\
Appreciation [\%] & 29.9 & 19.9 & 23.0 & 32.2 & 24.5 & 28.3 \\
\hline



 \end{tabularx}
\end{table}

\subsection{Approach}

In Table~\ref{tab:rq1eval}, we report numbers about the participants' evaluation of generated comments and, when applicable, their reasons for ignoring LLM-generated comments.
Even though participants were asked to evaluate all comments, we gave them the option to indicate cases where they felt unable to perform an evaluation (option ``Ignored - I'm not sure"). 
When considering all evaluated comments ($\#Evaluated$ in the ratios), we exclude those comments.

It should also be noted that due to differences in the review environment UI in the two organizations (see Section~\ref{sec:diffCompany}), reviewers from Ubisoft were not forced to evaluate the full set of generated comments per patch, but were encouraged to do so, contrary to Mozilla that enforced users to do full evaluations. 
This leads to comments that were seen but not evaluated in the case of Ubisoft, which we consider as ignored with no selected reason. We marked those as ``Ignore - Seen".
Acceptance and appreciation ratios are reported in Table~\ref{tab:rq1res}.



\subsection{Results} \label{rq1res}

\textbf{8.1\% and 7.2\% of generated comments were accepted, for Mozilla and Ubisoft respectively, as can be seen in Table~\ref{tab:rq1res}.}
For Ubisoft, the \textit{Example} approach reports a significantly better acceptance ratio (8.9\% vs 5.3\%) with a small effect size compared to the \textit{Code} approach (p-val=.011). For Mozilla, the \textit{Example} approach still reports a higher acceptance ratio (8.6\% vs 7.1\%), although the difference is not significant (p-val=.381). 
We suspect that the \textit{Example} approach has the advantage of showing comments more similar to what users are used to seeing and publishing themselves, as the few-shot examples shown by the \textit{Example} approach are taken from past similar published cases. 

\textbf{23\% and 28.3\% generated comments, comprising both accepted comments and valuable tips, are appreciated at Mozilla and Ubisoft respectively.}
Table~\ref{tab:rq1res} shows how, this time, the \textit{Code} approach reports a significantly higher appreciation (medium effect size) than the \textit{Example} approach, for both organizations (respectively, 29.9\% vs 19.9\% with p-val=.022 and 32.2\% vs 24.5\% with p-val=.002). 
This difference could indicate that the context related to the change provided by the \textit{Code} approach helps the model generate more informed and accurate comments. Yet, as the few-shot examples provided by this approach are quite limited compared to the \textit{Example} approach, and do not adapt to the precise style of comments required, the resulting comments do not seem to make it as publishable (accepted) review comments.


\textbf{Overall, both organizations report similar results, despite their differences in codebase availability, development and review processes.}
As indicated in Section~\ref{sec:companies}, Mozilla has the advantage that its codebase is open-source, with modern LLMs potentially having acquired knowledge about Mozilla's codebase through pre-training.
Yet, we do not observe any significant differences in results when comparing against each other both organizations, for any combination of ratios (i.e., acceptance and appreciation) and sets (i.e., Code, Example or Total) (p-value=.279 to .986).


\textbf{23/37 reviewers across both organizations reported that they would use a review assistive tool like \toolName\ at least sometimes outside of the study (question 5, Table~\ref{tab:survey}).} 
To analyze this closer, two authors aggregated the participants' mentioned concerns from the open answers to question 4 of the survey, which asked users what they would change about the comments' content. 
Overall, 20/37 participants explicitly mentioned that the relevancy of comments should be improved, as most of the suggested comments are ``\textit{accurate}", but ``\textit{superficial}" , ``\textit{obvious}", ``\textit{generic}", ``\textit{vague}", or sometimes even ``\textit{wrong}", as also confirmed by the numbers for ``Ignore - Trivial comment'' and ``Ignore - Incorrect comment'' in Table~\ref{tab:rq1eval}. 
Yet, one participant mentioned that ``\textit{the tool can be super useful for less experienced reviewers and a good way for them to learn}", confirmed by 2 other users.
Also, 6/37 mentioned being optimistic on the future of LLM-based approaches for review assistance tasks, and 11/37 mentioned that, at least, the tool was able to guide their review (i.e., providing a tip for reviewers).


\noindent\doublebox{%
    \parbox{.95\linewidth}{%
        \textbf{RQ1: Generated review comments achieve promising acceptance (8.1\% and 7.2\%) and appreciation ratios (23\% and 28.3\%). 23/37 reviewers that answered our survey reported that they would continue using \toolName~at least sometimes. 
        }
    }%
}

\begin{table}[t]
    \centering
    \caption{RQ2 - For each automatically generated comment category, we list the number of comments from each Organization (Moz and Ubi) that were labeled as such.``*" indicates that the label was an example in the few-shot prompt~\cite{turzo2024makes}. }
    \label{tab:cluster}
    \begin{tabularx}{\linewidth}{c|l|X|X}
    \hline
         \#&Label& \textbf{Moz.}& \textbf{Ubi.} \\
         \hline

1 & Discussion - Design discussion * & 3  & 3  \\
2 & Discussion - Question * & 7  & 1  \\
3 & Documentation * & 11  & 4  \\
4 & Functional - Conditional Compilation & 2  & 11  \\
5 & Functional - Consistency and Thread Safety & 1  & 0 \\
6 & Functional - Error Handling & 4  & 2  \\
7 & Functional - Exception Handling & 0 & 2  \\
8 & Functional - Initialization & 1  & 8  \\
9 & Functional - Interface *& 81  & 91  \\
10 & Functional - Lambda Usage & 2  & 3  \\
11 & Functional - Logical *& 74  & 272  \\
12 & Functional - Null Handling & 0 & 3  \\
13 & Functional - Performance & 3  & 18  \\
14 & Functional - Performance Optimization & 0 & 1  \\
15 & Functional - Performance and Safety & 0 & 1  \\
16 & Functional - Resource * & 58  & 135  \\
17 & Functional - Security & 1  & 0 \\
18 & Functional - Serialization & 0 & 4  \\
19 & Functional - Support * & 3  & 2  \\
20 & Functional - Syntax & 3  & 2  \\
21 & Functional - Timing * & 3  & 56  \\
22 & Functional - Type Safety & 0 & 2  \\
23 & Functional - Validation *& 95  & 392  \\
24 & Refactoring - Alternate Output * & 11  & 6  \\
25 & Refactoring - Code Duplication & 0 & 7  \\
26 & Refactoring - Code Simplification & 0 & 2  \\
27 & Refactoring - Consistency & 0 & 1  \\
28 & Refactoring - Magic Numbers & 0 & 1  \\
29 & Refactoring - Naming Convention * & 15  & 24  \\
30 & Refactoring - Organization of the code * & 10  & 79  \\
31 & Refactoring - Performance Optimization & 0 & 6  \\
32 & Refactoring - Readability & 3  & 11  \\
33 & Refactoring - Simplification & 0 & 3  \\
34 & Refactoring - Solution approach * & 26  & 27  \\
35 & Refactoring - Unused Variables & 0 & 1  \\
36 & Refactoring - Variable Declarations & 0 & 1  \\
37 & Refactoring - Visual Representation & 1  & 3  \\        
\hline
    \end{tabularx}
\end{table}




\begin{table}[t]
    \centering
    \caption{RQ2 - Categorization of generated comment and associated appreciation and acceptance.}
    \label{tab:categoryres}
    \begin{tabularx}{.99\linewidth}{l|XXXX|XXXX}
    \hline
        \textbf{Organization} & \multicolumn{4}{c|}{\textbf{Mozilla}}& \multicolumn{4}{c}{\textbf{Ubisoft}} \\
         \hline
    \diagbox[dir=NW, width=2.5cm, height=2cm]{\textbf{Metric}}{\textbf{Category}}&\rotatebox[origin=c]{75}{\textbf{Functional}}
    &\rotatebox[origin=c]{75}{\textbf{Refactoring}}
    &\rotatebox[origin=c]{75}{\textbf{Documentation}}
    &\rotatebox[origin=c]{75}{\textbf{Discussion}}
    &\rotatebox[origin=c]{75}{\textbf{Functional}}
    &\rotatebox[origin=c]{75}{\textbf{Refactoring}}
    &\rotatebox[origin=c]{75}{\textbf{Documentation}}
    &\rotatebox[origin=c]{75}{\textbf{Discussion}}\\
        \hline
        Count [\#]&331&66&11&10&1005&172&4&4\\
        \hline
        \%Total [\%]&79.2&15.8&2.6&2.4&84.8&14.5&.34&.34\\
        Acceptance [\%]&4.8&18.2&-&-&5.2&18.6&-&-\\
        Appreciation [\%]&19.0&36.4&-&-&29.0&25.0&-&-\\


\hline
    \end{tabularx}
\end{table}

\section{RQ2: \suggRQtwo}



\subsection{Approach}

To automatically categorize the generated comments, we clustered them according to their embeddings, then used an LLM to synthesize a category label for each cluster based on the comments that it contains\cite{labelcluster}. 
The categories we considered are the general categories presented by Turzo et al.~\cite{turzo2024makes}: functional, refactoring, documentation, and discussion.

First, comments are gathered into clusters using Kmeans\cite{kmean} on the embedded comment (using GPT's text-embedding-3-large).
Then, we ask the LLM to label each cluster of comments with a descriptive label, by providing it with a list of label examples of the form ``\textit{\{general\_category\} - \{specific\_category\}: \{description\}}'', corresponding to the table of categories presented by Turzo et al.~\cite{turzo2024makes}.
Given those examples, we observed that often, multiple clusters were labelled with the same label, which led us to eventually merge the 363/400 concerned clusters into the final 37 clusters shown in Table~\ref{tab:cluster}.
Also, not all generated labels existed in the original list of examples we provided, which we indicate by a star ``*" in Table~\ref{tab:cluster}.
We ran the LLM-base labelling 5 times and report for each comment the label that got the majority vote.

In order to determine the optimal number of clusters, we compared, for the number of clusters ranging from 2 to 400, the \textit{general\_category} of each cluster with that of a manually labelled ground truth, which we discuss below. We found that the percentage of matches with the ground truth converged between 300-400 clusters, leading to 76.3\% of matches (171/244) for 400 clusters, with only 37 distinct labels being generated even with that amount of clusters. 
To evaluate the agreement between the model and the ground truth (human labelling), we used Cohen's kappa score, a metric between -1 and +1 where +1 indicates total agreement and -1 total disagreement~\cite{cohen1960coefficient}. 
We obtained a kappa score of .45, which is considered as moderate but acceptable agreement.

To obtain the manually labelled ground truth, used to determine the optimal number of clusters, we sampled from each of the macro-evaluation categories of RQ1 (i.e., accepted, valuable tip, and rejected) 83 generated comments (70\% of \#accepted, the ones available at the time of the analysis), in order to have a balanced set of each type, for a total of 249 comments.
Then, two authors manually labelled each sample with a comment category using Turzo et al.'s main categories~\cite{turzo2024makes}.
We did not consider their ``false positive'' category, as we did not evaluate the validity of the comment, only its category. 
In the end, in 93\% of the cases (230/249) the authors agreed on the selection of one out of the four possible categories.
The kappa score between both authors' labelling of .83 is considered as almost perfect agreement.
For any conflicts, the two authors performing the labelling discussed and eventually agreed on the category using the \textit{negotiating agreement} technique~\cite{campbell2013coding}, but 5 comments were left unlabelled because of ambiguity in the comments themselves.
The latter were discarded, leaving us with 244 labelled comments.





For the selected number of clusters (400) and resulting cluster labels, we now split comments into groups according to the generated labels' \textit{general\_category}, shown in Table~\ref{tab:categoryres}, as well as their acceptance and appreciation ratios. 

\subsection{Results}

\textbf{Table~\ref{tab:categoryres} shows that Functional and Refactoring comments are the most commonly generated comments, with respectively 79.2\% and 15.8\% for Mozilla and 84.8\% and 14.5\% for Ubisoft.} 
Notably, Table~\ref{tab:cluster} shows that the most popular sub-categories of Functional comments are ``Validation'', ``Logical", ``Resource" and ``Interface", and ``Organization of the code" and ``Solution approach" in the case of Refactoring comments. Those subcategories were proposed through the few-shot examples provided to the LLM, from Turzo et al.~\cite{turzo2024makes}.
The Documentation and Discussion labels are the rarest, with a total of 21/418 comments for Mozilla and 8/1,185 for Ubisoft.
This is expected, as the prompt ``\textit{AskLLMReview}'' (see prompt (5) in Table~\ref{tab:prompts}) suggests that such comments are not desirable, which is also why we did not report their acceptance and appreciation ratios in Table~\ref{tab:categoryres}.

\textbf{Refactoring comments are accepted more than functional comments (18.2\% vs 4.8\% for Mozilla and 18.6\% vs 5.2\% for Ubisoft).}
This difference is statistically significant with large effect size, with respectively p-values .0 and .002. 

\textbf{In Mozilla, appreciation for refactoring comments is significantly higher than for functional comments (36.4\% vs 19\%, p-val=.002, large effect size).}
In the case of Ubisoft, the difference is not significant (p-val=.877), although in practice Functional comments are more appreciated (29\% vs 25\%).

\noindent\doublebox{%
    \parbox{.95\linewidth}{%
        \textbf{RQ2: Refactoring comments report a significantly higher acceptance ratio than Functional comments in most cases (18.2 vs 4.8\% for Mozilla and 18.6 vs 5.2\% for Ubisoft). Discussion and Documentation comments are rare (29/1,603 comments), which is expected based on our prompt design.
        }
    }%
}

\section{RQ3: \suggRQthree}



\subsection{Approach}

\begin{figure}[t]
    \centering
    \includegraphics[width=\linewidth]{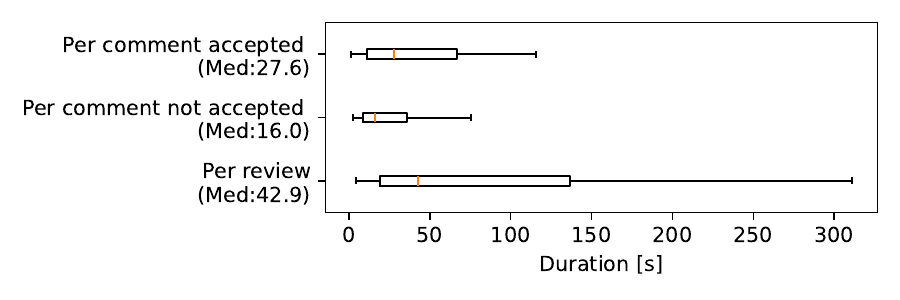}
    \caption{RQ3 - Evaluation duration. Outliers are hidden for visibility. }
    \label{fig:RQ4_duration}
\end{figure}

In question 3 of the survey (Table~\ref{tab:survey}), we asked participants whether interacting with the generated comments impacted the duration of their reviews. 
In the case of Ubisoft, we also captured the time it took to evaluate the generated comments in order to empirically estimate that time.
This was possible due to the difference in design (see Section~\ref{sec:diffCompany}): for each generated comment, we marked the timestamp of when a given comment was first opened (not applicable to Mozilla) and of the timestamp when the reviewer submitted their evaluation of this comment. The difference between those timestamps provides an upper bound on the actual time it took a reviewer to evaluate a generated comment, as this period is longer than the actual time taken for evaluation due to switching to other comments or distractions.
In Figure~\ref{fig:RQ4_duration}'s boxplots, we report the distribution of the duration of evaluation per comment and per patch review (i.e., total duration for all its generated comments' evaluations).

We also evaluate how reviewers edit the accepted review comments. In Table~\ref{tab:edits}, we report four different types of edits on accepted comments:
\begin{itemize}
    \item As-is: with no modification (i.e., exact match between generated and published comment);
    \item Shorten: the published comment is shorter than the generated one;
    \item Extended: the generated comment is included as-is in the published one, but more text was added around it;
    \item Other modification: any modification not included above.
\end{itemize} 

\subsection{Results}

\begin{table}[t]
    \centering
    \caption{RQ3 - Editing of accepted comments}
    \label{tab:edits}
    \begin{tabularx}{\linewidth}{l|X|X}
    \hline
         \textbf{Evaluation}& \multicolumn{1}{c|}{\textbf{Organization1}}& \multicolumn{1}{c}{\textbf{Organization2}} \\
         \hline
Suggestion as-is&    27  & 55\\
Shortened content&    6 & 19\\
Additional content&  1  & 4\\
Other modification&  0 & 7\\
         \hline 
    \end{tabularx}
\end{table}


\textbf{The median time spent at Ubisoft on the evaluation of individual comments is 27.6s for accepted comments and 16s for others, as shown in Figure~\ref{fig:RQ4_duration}.} 
By 24/37 users in the survey, the additional effort of evaluating comments was reported as, to some extent, slowing down the review process. 
Our quantitative analysis of the comment timestamps showed that this slowdown is relatively limited, falling within a 95\% confidence interval of [1s,116s] (median of 27.6s) for accepted comments, and [1s,76s] (median of 16s) for others.
Although time spent on ignored comments might be considered ``wasted time'' by reviewers, in some cases the comments might still inspire the reviewer to comment on the code about problems they would have missed otherwise, or at least force an extra verification of the code. We did not find any significant difference in the median time spent evaluating appreciated comments vs fully rejected comments, hence did not include the former in Fig.~\ref{fig:RQ4_duration}.

\textbf{By patch, the median total time spent on generated comments is 42.9sec (95\% confidence interval [5s,5min12s]).}
This effort is considered by many users as acceptable. For instance, one of the reviewers commented: \textit{``Often looking at the suggestion helps me think more about the code and reflect a bit more on what is done and why it is done. So overall, I think it is a bit slower to do reviews, but my reviews are actually better now that I use the tool"}.
Similar comments were made by 11/37 users.
The extra time spent on the generated comments also comes from either verifying the validity of the comment or, specifically for the accepted ones, modifying the comments.

\textbf{27/34 accepted comments were published as-is for Mozilla, 55/85 for Ubisoft, as shown in Table~\ref{tab:edits}}
The most common type of edit to generated comments is to shorten the comment (6/34 for Mozilla, 19/85 for Ubisoft), which suggests that generated comments can often be too wordy, which was confirmed in the survey by three participants. 
In future work, we want to iterate on the prompt to make generated comments more succinct.

\noindent\doublebox{%
    \parbox{.95\linewidth}{%
        \textbf{RQ3: 24/37 surveyed reviewers report that review duration is increased by our tool. We quantitatively measured that the median time dedicated to the inspection of generated comments at the review level is 43s in Ubisoft, which users find reasonable. In both organizations, most comments, when accepted, are mostly published as-is (82/119) or shortened (25/119).
        }
    }%
}


\section{RQ4: \suggRQfour}



\subsection{Approach}

To evaluate the impact of accepted comments on the further processing of the patch, we captured, for each line commented on via an accepted generated comment, whether that comment led to threads of review comments and/or follow-up revisions. 

A potential impact of the comment would be any follow-up review comments, i.e., threads of comments, made by the author of the patch or other reviewers in response to the posted comment to (1) better understand the needed change, (2) argue about the need for adjusted revisions suggested or (3) acknowledge the actions taken for the requested change.
It should be noted that the follow-up revisions might not be directly related to the accepted comments, but at least this characteristic provides a reasonable approximation of the potential impact the comment may have had on the evaluation of the commented patch, in both generated comments and human comments from patch reviews outside of the user study. 

For the analysis of the impact of generated comments on later patch revisions and/or review discussions, we selected comments based on the following constraints.
First, we only consider accepted comments.
Then, we restricted the selection to patches with ``finished'' review status, since some reviews were still ongoing at the time of the study, thus the comment might not yet have been dealt with.
Finally, we considered only results from Ubisoft, which represent 69 comments satisfying the selection criteria, as Mozilla had only 25 cases after filtering, which would not have led to significant comparison. 

As a baseline for our analysis on generated comments, we also collected a dataset of ``Human" comments, by other reviewers not part of the user study, sampled randomly from patches under review in Ubisoft but that were not included in the actual user study, yielding a total of 1,211 comments.




We report in Table~\ref{tab:rq4res} three different types of impact that accepted comments can exhibit:

\begin{itemize}
    \item \textit{Revised line}: The comment's line was modified in a follow-up patch revision.

\item \textit{Revised chunk}: The comment's line was modified or was part of the context in a follow-up patch revision.

\item \textit{Thread}: There was a comment in response to it. 
\end{itemize}

\subsection{Results}

\textbf{In Ubisoft, generated comments lead to as many revisions on the line (62.3\% and 64.3\%) and chunk level (73.9\% and 73.2\%) as human comments do.} 
No significant difference is found in revision ratios on either levels (p-val=.413 and p-val=.515, respectively). 

\textbf{For Ubisoft, significantly less comment threads were started on generated comments (23.2\%) compared to human comments (33.9\%).}
The difference is statistically significant (p-val=.041) with medium effect size.
This is expected, as the generated comments are not meant to start discussions, whereas human comments can often, for instance, be phrased as genuine questions requesting a response.  
This expands on our findings from RQ2 that most comments are not of the Discussions category.
However, from manual inspection of the concerned comments, in practice the threads tend to discuss the generated comment and/or to acknowledge the change (i.e., writing ``Done'') or reject it.

\begin{table}[t]
    \centering
    \caption{RQ4 - Impact of generated comments on finished patch reviews. ``Generated" refers to accepted generated comment. ``Human" refers to manually written comments on patches without generated comments.}
    \label{tab:rq4res}
    \begin{tabularx}{\linewidth}{r|Xr|Xr}
    \hline
         \textbf{Evaluation}& \multicolumn{4}{c}{\textbf{Ubisoft}} \\
         & \multicolumn{2}{c|}{\textbf{Generated}}& \multicolumn{2}{c}{\textbf{Human}} \\
         \hline

Revised line & 43 &62.3\% & 779 &64.3\% \\
Revised hunk & 51 &73.9\% & 887 &73.2\% \\
Thread  & 16 &23.2\% & 411 &33.9\% \\
\hline Total  & 69 && 1211 \\
\hline
    \end{tabularx}
\end{table}

\noindent\doublebox{%
    \parbox{.95\linewidth}{%
        \textbf{RQ4: Generated comments experienced as many revisions as human comments (73.9 vs 73.2\% at chunk-level), but less follow-up comments (23.2 vs 33.9\%). 
        }
    }%
}

\section{Discussion}
\label{sec:discussion}

Evaluating review comment generation is a challenging task, especially as it is essential to capture the impact of assisting reviewers in their work.
Where prior work has addressed the question of review comment generation by textually comparing the generated outputs with a ground truth of past chunks commented on by humans, such metrics do not evaluate the impact the generation has on the reviewers' workflow.
In this paper, we instead carried out a user study to gather feedback from reviewers in two software development organizations. 

The first metric we report is acceptance, i.e., the ratio at which reviewers actually accepted comments for publication. 
We observed that reviewers, even if they ignore the comments, often still report them as having value (14.6\% and 20.5\%).

First, reviewers reported that 5.4\% and 12.8\% comments, respectively, could actually be more valuable for developers before they submit their patch. Those generated comments are thus shown too late in the development process.
One participant reported in the survey ``\textit{It detects perfectly possible flaws in the code, but most of the suggestions are more in the development realm than the review}", with 4 participants making similar comments. That reason for ignoring LLM-generated comments was used by 28/57 reviewers. 

Second, although some comments are not actionable as-is or not mentioning specific problems, they often (9.2\% and 7.7\%) raise interesting questions ``\textit{that force to look around, and it is what is found by looking around that is the most useful}", as mentioned by one participant. Another even suggests, ``\textit{there's more value in highlighting potential issues with a more concise explanation and let the reviewer write the comment}", and similar comments were provided by 10 other participants. The ``review tip" reason was used by 25/57 participants.

The high ratio of appreciation (23\% and 28.3\%) compared to comment acceptance highlights that the comment generation approach actually addresses multiple tasks: suggesting actionable comments, developer tips and review tips.
Future work should explore the possibility to split the overall review comment generation task by predicting the type of generated comments or specializing towards the generation of specific comment types. Developer tips should be shown ahead of the review process (perhaps giving developers access to tools like \toolName\ before patch submission), and review tips should be shown to reviewers differently than actionable comments, as they act as a warning for reviewers to focus on some specific code snippet.
The same observation can be done regarding categories of comments: future work could explore the categories of review comments, considering generation of each category as a separate task, and focus the generation on specific categories, depending on reviewers' preferences.

\section{Threats to Validity}
\label{sec:threats}

\emph{\textbf{Construct to Validity:}} 
\toolName\ currently is limited to two RAG approaches, one retrieving extra code context, the other relevant examples of past chunks with their review comment. 
Although other approaches could be explored, we chose those two
to cover a wider range of modern LLM techniques currently applied in the SE domain~\cite{zhou2022docprompting, parvez2021retrieval}.

\emph{\textbf{Internal Validity:}} 
Reviewers might accept or reject comments based on their personal judgment and/or style of reviews. 
When evaluating productivity, different factors, like deadlines, policies, preferences, and tool changes, might affect our results, which we cannot control during our investigation. 

When classifying the categories of generated comments in RQ2, we perform an automated labelling using LLM.
To mitigate the potential hallucination of the model, we compared the automated labelling with a ground truth build through manual labelling.
The LLM was provided with a few-shot list of categories~\cite{turzo2024makes}. 
When compared with the ground truth, the automated labelling reached a moderate agreement with the ground truth, i.e., we observed no substantial hallucination.

\emph{\textbf{External Validity:}} Exploring review comment generation for other organizations might impact results based on their review policies and practices.
However, our investigation covers two organizations, one being open-source, and one closed-source. Despite their differences, we found that the two organizations achieved similar results in the user study.

\toolName\ generates comments based on OpenAI's GPT4o.
LLMs are known to be computationally expensive, regarding both training and generations.
The approach we consider uses an off-the-shelf model with few-shot and chain-of-thoughts prompt architecture, thus avoiding the additional costs of fine-tuning the model and maintaining it.
Future work should consider exploring fine-tuning the model, 
regarding the trade-off between the improvement on comment generation and the related extra costs (e.g., computational effort). 



\section{Conclusion}
\label{sec:conclusion}

Code review plays an essential role in modern collaborative software development, representing a way to improve the quality of evolving systems while improving developers' social and technical skills~\cite{mcintosh2016empirical}. 
Recent studies have introduced review comment generation tasks into the review process, enhanced by LLM-based approaches~\cite{tufano2022using, li2022automating}.
However, they have not evaluated the impact of such approaches on the review process.

Through a large-scale mixed open-/closed-source user study, we report acceptance ratios of 8.1\% for Mozilla and 7.2\% for Ubisoft, and we further observe that appreciation is higher (23\% and 28.3\%, respectively). 
Refactoring comments, despite being the second most popular type of generated comments (15.8\% and 14.5\%) after functional comments (79.2\% and 84.8\%), have significantly better acceptance levels (18.2\% vs 4.8\% and 18.6\% vs 5.2\%).
The impact of the LLM-based approach on the review workflow is reasonable as reviewers spend a median time of 43s per review on investigating generated comments and, when applicable, editing accepted ones (37/119) to, in most cases, shorten them (25/37).
Regarding accepted comments, we find that their impact on patches’ review processes is similar to human-written comments, as 74\% vs 73\% of comments have at least one follow-up revision at chunk-level, which is promising for future adoption of LLM-generated review comments.

We hypothesize that the gap between acceptance and appreciation of generated review comments stems from the task not being properly defined for the LLM. 
Indeed, the LLM involved addresses multiple tasks, i.e., generating publishable comments, tips for developers and tips for reviewers, all of those being valuable to the review process.
We conclude that the current way of generating and presenting the results of review comment generation to reviewers and patch authors in the review environment should be reconsidered.
\bibliographystyle{IEEEtran}
\bibliography{sample-base}


\end{document}